\documentstyle[11pt,moriond2,epsfig]{article}

\def\Journal#1#2#3#4{{#1} {\bf #2}, #3 (#4)}

\def\NPB{{\em Nucl. Phys.} B}
\def\PLB{{\em Phys. Lett.}  B}
\def\PRL{\em Phys. Rev. Lett.}
\def\PRD{{\em Phys. Rev.} D}

\def\be{\begin{equation}}
\def\ee{\end{equation}}
\def\bea{\begin{eqnarray}}
\def\eea{\end{eqnarray}}

\def\simgt{\stackrel{>}{{}_\sim}}
\begin{document}
\begin{flushright}
        {\normalsize CPHT-PC715.0499}\\
        {\normalsize hep-ph/9904272}\\
\end{flushright}

\vspace*{1cm}
\title{ON POSSIBLE MODIFICATIONS OF GRAVITATION IN THE (SUB)MILLIMETER RANGE}

\author{ I. ANTONIADIS }

\address{Centre de Physique Th\'eorique (CNRS UMR 7644), Ecole Polytechnique,\\
91128 Palaiseau Cedex, France}

\maketitle\abstracts{
I discuss possible modifications of gravitation at short distances in string 
theories with large internal dimensions and low string scale. The modifications
are due to the change of Newton's law in the presence of (sub)millimeter-size
transverse dimensions, or due to the existence of light scalars with masses as 
small as 10$^{-3}$ eV, related to the mechanism of supersymmetry breaking,
mediating new Yukawa-type forces with strength comparable to gravity. These
modifications  are testable in ``tabletop" experiments that measure gravity at
such short distances.\\
(Invited talk given at the XXXIV Rencontres de Moriond on 
{\em Gravitational Waves and Experimental Gravity}, 
Les Arcs, France, January 23-30, 1999)}

\section{TeV$^{-1}$ dimensions, supersymmetry breaking and submm forces}

String theories provide at present the only consistent framework of quantum gravity
and unification of all fundamental interactions. Traditionally, it was believed
that they become relevant only at Planckian energies due to the simple tree-level
expression for the string scale, 
\be
M_H\simeq gM_P\, ,
\label{het}
\ee
valid in the context of perturbative heterotic theory. 
Here, $M_P=1.2\times 10^{19}$ GeV is the Planck
mass, and $g\simeq 1/5$ is the gauge coupling at the heterotic string (unification)
scale $M_H\simeq 10^{18}$ GeV. Despite this relation, there are however physical
motivations which indicate that large volume compactifications may be relevant for
physics. One of them comes from the problem of supersymmetry breaking by 
the process of compactification which relates the breaking scale to the size of
some internal compact dimension(s). The latter should therefore be in the
TeV$^{-1}$ region in order to keep the scale of supersymmetry breaking close to
electroweak energies~\cite{a}.

An immediate defect of this scenario is that the ten-dimensional (10d) string
coupling is huge invalidating the perturbative description. From the 4d point of
view, the problem arises due to the infinite massive tower of Kaluza-Klein (KK)
excitations that are produced at energies above the compactification scale and
contribute to physical processes: 
\be
M^2=M_0^2+{n^2\over R^2}\ ;\qquad n=0,\pm 1,\dots\, ,
\label{KK}
\ee
where $M_0$ is a 5d mass and $R$ is the radius of an extra dimension.
If the theory makes sense for some special
models, a possible way out consists of imposing a set of conditions to the low
energy theory that prevent the effective couplings to diverge. An example of such
conditions is that KK modes should be organized into multiplets of $N=4$
supersymmetry, containing for each spin-1 4 Weyl-fermions and 6 scalars with the
same quantum numbers, so that their contribution to beta-functions vanish for every
$n\neq 0$ and gauge couplings remain finite.

The breaking of supersymmetry by compactification is realized through boundary
conditions and is similar to the effects of finite temperature with the
identification $T\equiv R^{-1}$. It follows that the breaking is extremely
soft and insensitive to the ultraviolet (UV) physics above the compactification
scale. The summation over the KK excitations amounts to inserting the Boltzmann
factors $e^{-E/T}$ to all thermodynamic quantities --or equivalently to the soft
breaking terms--  that suppress exponentially their UV behavior~\cite{a,add}.
This is in contrast to the behavior of supersymmetric couplings that generally
blow up, unless special conditions are imposed. 

The extreme softness of supersymmetry breaking by compactification
has two important phenomenological consequences: (i) a particular spectroscopy of
superparticles that differs drastically from other scenaria~\cite{a,adpq}, and
(ii) a new scalar force in the (sub)mm range mediated by the scalar modulus whose
vacuum expectation value (VEV) determines the radius $R$ of the TeV dimension
used to break supersymmetry~\cite{add}. One can easily check by direct dimensional
reduction of the higher-dimensional Einstein action that the scalar field associated
to the radius modulus with canonical kinetic terms is $\phi=\ln R$ in Planck units.
Its VEV corresponds to a flat direction of the scalar potential, which is lifted
only after supersymmetry breaking, through a potential generated by radiative
corrections. The softness of the breaking mechanism implies that the vacuum energy
$V$ behaves for large $R$ as
\be
V\sim T^4\equiv{1\over R^4}\, ,
\label{vacen}
\ee
up to logarithmic higher loop corrections. Notice the absence of a
``quadratically divergent" $M_P^2 R^{-2}$ contribution. It follows that the mass
of the radius modulus $\phi$ is of order~\cite{fkz,add}
\be
m_\phi\sim R^{-2}/M_P\simeq 10^{-4} - 10^{-2}\ {\rm eV} \, ,
\label{mphi}
\ee
for a compactification scale $R^{-1}\simeq 1 - 10$ TeV. This corresponds to
Compton wavelengths of the modulus in the range of 1 mm to 10 $\mu$m.

The coupling of this light scalar to matter arises dominantly through the
dependence on $\phi$ of the QCD scale $\Lambda_{\rm QCD}$, or equivalently of
the QCD gauge coupling $\alpha_{\rm QCD}$. The coupling of the radius modulus to
nucleons is given by the derivative of the nucleon mass relative to $\phi$,
${\partial m_N\over\partial\phi}$. Since the graviton coupling to nucleons is
$m_N/M_P$, the coupling $\alpha_\phi$ of the modulus relative to gravity is
\be
\alpha_\phi = {1\over m_N}{\partial m_N\over\partial\phi}=
{\partial\ln\Lambda_{\rm QCD}\over\partial\ln R}=
-{1\over b_{\rm QCD}}{\partial\over\partial\ln R}{2\pi\over\alpha_{\rm QCD}}\, ,
\label{coupling}
\ee
with $b_{\rm QCD}$ the one-loop QCD beta-function coefficient.
Since in these models, in the absence of supersymmetry breaking all couplings of
the effective field theory are well behaved in the large radius limit, while after
supersymmetry breaking all soft masses are proportional to $1/R$, one has 
\be
{\partial\over\partial\ln R}{2\pi\over\alpha_{\rm QCD}}=
b_{\rm SQCD}-b_{\rm QCD}\, ,
\label{deltaR}
\ee
with $b_{\rm SQCD}$ the supersymmetric beta-function. Using $b_{\rm SQCD}=3$ and
$b_{\rm QCD}=7$, one obtains $\alpha_\phi=1-b_{\rm SQCD}/b_{\rm QCD}=4/7$. Thus,
the force between two pieces of matter mediated by the radius modulus is
$(4/7)^2 \simeq 1/3$ times the force of gravity~\cite{add}.

In principle there can be other light moduli which couple with larger  
strengths. For example the dilaton, whose VEV determines the (logarithm of the)
string coupling constant, if it does not acquire large mass from some dynamical
supersymmetric mechanism, leads to the strongest effect.
Its coupling is $\alpha_{\rm dilaton}=\ln(M_P/\Lambda_{\rm QCD})\sim 44$~\cite{tv},
which corresponds to a strength $\sim 2000$ times bigger than gravity. 

\begin{figure}
\centerline{\psfig{figure=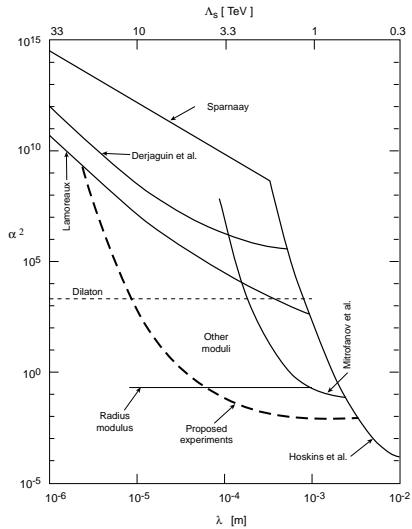,height=7cm}}
\caption{Strength of the modulus force relative to gravity ($\alpha^2$) versus 
its Compton wavelength ($\lambda$).
\label{fig:forces}}
\end{figure}

In fig.~\ref{fig:forces} we depict the theoretical
predictions together with information  from previous, present and upcoming
experiments. The vertical axis is the strength, 
$\alpha^2$, of the force relative to gravity; the horizontal axis is the 
Compton wavelength of the exchanged particle; the upper scale shows the
corresponding value of the large radius in TeV.
The solid lines indicate the present limits from 
the experiments indicated~\cite{oldexperiments}. The excluded regions lie 
above these solid lines. Our theoretical prediction for the radius modulus is 
the thin horizontal line at 1/3 times gravity. The dilaton-mediated force is 
the thin dashed horizontal line at $\sim$ 2000 times gravity. Other moduli may
mediate forces between the above two extremes.

Of course, measuring gravitational strength forces at such short  
distances is quite challenging. The most important background is  
the Van der Walls force. The Van der Walls
and gravitational forces between two atoms are equal to each other  
when the atoms are about 100 $\mu$m apart. Since the Van der Walls  
force falls off as the 7th power of the distance, it rapidly becomes  
negligible compared to gravity at distances exceeding 100 $\mu$m.
The most important part of the figure is the dashed thick line; it is the 
expected sensitivity of the present and upcoming experiments, which
will improve the actual limits by almost 6 orders of magnitude and --at the 
very least-- they will, for the first time, measure gravity to a precision of 
1\% at distances of $\sim$ 100 $\mu$m~\cite{price}.

\section{TeV strings, submm dimensions and signals in particle colliders}

As mentioned before, the main defect of the above heterotic string models
is the strong 10d coupling due to the large internal dimension(s) used to break
supersymmetry. Fortunately, this problem can be addressed using the recent results
on string dualities, by considering the corresponding dual weakly coupled theory.
In the minimal case of one or two large dimensions, the dual weakly coupled
description is provided by type II string theory, whose tension appears as a
non-perturbative threshold below the heterotic scale~\cite{ap}. For instance, in
the case of one TeV dimension, the type II scale appears at intermediate energies
of order $10^{11}$ GeV, and the theory above the TeV but below the intermediate
scale  becomes effectively six-dimensional and is described by a tensionless
string~\cite{ts} with no gravity. This 6d theory possesses a non-trivial infrared
fixed point which encodes the conditions needed to be imposed in the low energy
couplings, in order to ensure a smooth UV behavior. These conditions
generalize the requirement of $N=4$ supersymmetry for the KK excitations, that
keep the gauge couplings well behaved, to the Yukawa couplings of the theory.
A generic property of these models is that chiral matter is localized at particular
points of the large internal dimensions. As a result, quarks and leptons have not
KK TeV excitations, which is welcome also for phenomenological reasons in order to
avoid fast proton decay~\cite{a,ab}.

The main prediction of these theories in particle accelerators is the existence of
KK excitations~(\ref{KK}) for all Standard Model gauge bosons that can be produced
for instance in hadron colliders, such as the Tevatron and LHC, through Drell-Yan 
processes~\cite{abq}. The corresponding KK resonances are narrow with a ratio of
their width to their mass $\Gamma/m\sim g^2\sim$ a few per cent, and thus the
typical expected signal is the production of a double resonance corresponding
the first KK mode of the photon and $Z$, very nearly spaced the one from the other.
The current limits on the size of large dimensions arise from the bounds of
compositeness or from other indirect effects, such as in the Fermi constant and
LEP2 data, and lie in the range of $1-2$ TeV~\cite{ab,limits}. On the other hand,
direct production of KK states in LHC is possible for compactification scales of
less than about 5 TeV.

In the case of two TeV dimensions, the type II string scale appears also at the
TeV, while its coupling is infinitesimally small $\sim 10^{-14}$~\cite{ap}.
As a result, despite the fact that the string scale of the dual theory is so low,
it has no observable effects in particle accelerators and the main experimental
signal continues to be the existence of KK excitations along the two TeV
dimensions with gauge interactions. The predictions are thus very similar to the
previous case described above. It is amazing that the main features of these models
were captured already in the context of heterotic string despite its strong
coupling. When there are more than two large dimensions, a new phenomenon appears.
The dual weakly coupled theory, that can now be either type II or type I, contains 
also some extra large dimensions of size much larger than TeV$^{-1}\simeq 10^{-16}$
cm, where only gravity propagates.

It is now clear from the above discussion that in other than the heterotic 
perturbative descriptions of string theory,
the string scale is not tied to the Planck mass by a
simple relation like~(\ref{het}), but it corresponds to an arbitrary parameter
that can be everywhere above the TeV scale~\cite{wl}. In particular, if it is at
the TeV, the gauge hierarchy problem would be automatically solved without the
need of low energy supersymmetry or technicolor. The price to
pay is to introduce an extremely small string coupling and/or, alternatively,
extra large dimensions in order to account for the hierarchy $M_{\rm str}/M_P$.

For instance, it was recently
proposed that the fundamental scale of gravity could be at the TeV if there were
such extra large dimensions where only gravity propagates, while the Standard
Model should be localized on a 3d wall~\cite{add1,aadd}. The observed weakness of 4d
gravity would then be a consequence of the higher dimensional Gauss-law:
\be
G_N={G_{N(4+p)}\over R_{\perp}^p}\, ,
\label{gl}
\ee
where $G_{N(4+p)}$ is the $(4+p)$-dimensional Newton's constant and $R_{\perp}$
denotes the (common) size of $p$ extra large spacial dimensions. 
This scenario can be realized naturally in the context of type I theory of closed
and open strings~\cite{aadd,stk}. Gravity is described by closed strings, while
gauge interactions are described by open strings whose ends are confined to
propagate on subspaces, called D-branes. The theory is weakly coupled when the
size of the internal dimensions parallel to the D-brane where the Standard Model
is localized is of order of the string scale $M_I^{-1}$, while there is no
restriction on the size of transverse dimensions. One then obtains that the type I
string scale is simply related to the higher-dimensional Newton's constant
\be
G_{N(4+p)}\simeq g^2 M_I^{-2-p}\, ,
\label{typeI}
\ee
and is therefore a free parameter, while the string coupling is given by
the square of the gauge coupling. This expression should be compared with the
corresponding relation~(\ref{het}) in the heterotic string, where the string
scale is fixed by the 4d Planck mass.\footnote{Note that in the context
of type II theories, the string tension is decoupled from the scale where
gravity becomes strong and can be lowered independently by decreasing the string
coupling which is a free parameter.}

Using eqs.~(\ref{gl}) and~(\ref{typeI}) and
taking the type I string scale to be at the TeV, one finds a size for the
transverse dimensions varying from $10^8$ km, .1 mm, up to .1 fermi for $p=1,2$,
or 6 large dimensions. With the exception of $p=1$ which is obviously
experimentally excluded, all other values are consistent with
observations~\cite{add1,add2,ns}. The strongest bound applies for $p=2$ and comes from
graviton emission on cooling of supernovae which restricts the 6d Planck scale to be
larger than about 50 TeV~\cite{sn}, implying $M_I\simgt 7$ TeV. The main distinct
experimental signals of these theories in particle accelerators
are~\cite{aadd}:\hfil\\  
(i) Production of higher spin Regge-excitations for all
Standard Model particles with the same quantum numbers and mass-squared increasing
linearly with the spin. For instance, the excitations of the gluon could show up as
a series of peaks in jet production at LHC. However, the corresponding resonances
are very narrow, with a ratio of their width to their mass $\Gamma/m\sim g^4\sim$ a
few per thousand, and thus are difficult to detect.\hfil\\
(ii) Graviton emission into the extra dimensions leading to events of jets + missing
energy. LHC is sensitive to higher-dimensional gravity scales in the range of
3 to 5 TeV, when the number of transverse dimensions varies from $p=6$ at the 
subfermi to $p=2$ at the submm, where the effect becomes stronger~\cite{ns,grav}. This
is comparable to the sensitivity on possible TeV dimensions with gauge
interactions, as discussed previously. The most exciting possibility is of
course when the available energy is bigger than the gravity scale, in which
case gravitational interactions are strong and particle colliders become the best
tools for studying quantum gravity.

When the string scale is of order of a few TeV, the question of gauge hierarchy is
to understand why some dimensions transverse to our brane-world are so large
or/and why the string coupling is so small. Thus, the large number to explain
varies from $10^{15}$ up to $10^5$ in the case of six transverse dimensions.
This problem has a difficult part which
is its dynamical aspect and a {\it technical} part that consists to guarantee its
stability under quantum corrections. Supersymmetry solves this technical aspect by
allowing only logarithmic dependence on the UV cutoff --usually taken at $M_P$--
to the masses and other parameters of the Standard Model. Although naively one
would expect that low string scale would also solve the problem by
nullification~\cite{add1}, it turns out that this is not true in general~\cite{aba}.
The reason is due to the appearance of local tadpoles in various processes on the
world-brane, associated to the emission of a massless closed string in the
transverse space. These tadpoles diverge linearly when the closed string propagates
``effectively" in one preferred large transverse direction ($d_\perp=1$)
and logarithmically when $d_\perp=2$. Local tadpole cancellation implies severe
constraints for model building while their presence implies a set of conditions for
the compactification manifold that forbids linear divergences. The remaining
divergences are at most logarithmic and the technical aspect of gauge hierarchy is
solved at the same level as with supersymmetry. Moreover, the case of $d_\perp=2$ is
singled out as the only one in which the origin of the hierarchy would not be
attributed to `out of this world' bulk physics, since the presence of logarithmic
sensitivity could allow for a dynamical determination of the hierarchy by minimizing
the effective potential.\footnote{A similar conclusion appears to hold in low-scale
type II models with infinitesimal coupling~\cite{ap}.}

Although the initial motivation of low energy supersymmetry is in principle lost,
there may be still a reason to be introduced in the bulk, in connection to the
cosmological constant and stability problem. In fact, the bulk energy density of a
generic non-supersymmetric string model is $V_{\rm bulk}\sim M_{\rm str}^{4+p}$,
giving rise to an energy density on our world-brane~\cite{adpq}:
\be
V_{\rm brane}\simeq V_{\rm bulk}R_{\perp}^p\sim M_{\rm str}^2 M_P^2\, ,
\label{rho}
\ee
by virtue of eqs.~(\ref{gl}) and~(\ref{typeI}). This is analogous to the
quadratically divergent contribution to the vacuum energy in softly broken
supersymmetry, which in general destabilizes the hierarchy since it
induces a large potential for $R_\perp$. On the other hand, its absence is
guaranteed if the bulk remains approximately supersymmetric, which is the case
when for instance supersymmetry is broken spontaneously primordially only on the
brane (even maximally)~\cite{aadd}.

\section{Gravity modification at short distances}

TeV scale strings with extra large transverse dimensions predict important
modifications of gravitation at short distances that could be measured in
experiments which test gravity in the submillimeter range (see
fig.~\ref{fig:forces}). These ``tabletop" experiments, that cost orders of
magnitude less than high energy particle accelerators, might provide an independent
verification of the above ideas much before LHC. In fact, there are two classes of
general predictions:\hfil\\ (i) Deviations from the Newton's law $1/r^2$ behavior
to $1/r^{2+p}$ which can be observable for $p=2$ transverse dimensions of submm
size. This case is particularly favored for theoretical reasons because of the
logarithmic sensitivity of Standard Model couplings on the size of transverse
space, as we discussed above, but also for phenomenological reasons, since the
effects to particle colliders are maximally enhanced. Notice also the coincidence
of this scale with the possible value of the cosmological constant in the universe
that recent observations seem to support.\hfil\\
(ii) New scalar forces in the submm range mediated by light bulk scalar fields with
masses $M_{\rm str}^2/M_P$, motivated by the problem of supersymmetry breaking.
Such a universal scalar corresponds to the transverse radius
modulus $\phi_\perp\equiv\ln R_\perp$ in Planck units. Following our previous
discussion, if supersymmetry is broken primordially only on the brane, its
transmission to the bulk is gravitationally suppressed and $\phi_\perp$ acquires a
mass proportional to the gravitino mass, of the order of~\cite{aadd}
\be
m^2_{\phi_\perp}\sim G_{N(4+p)}{V_{\rm brane}\over R_{\perp}^p}\simeq
{M_{\rm str}^4\over M_P^2}\, ,
\label{mphiperp}
\ee
using eq.~(\ref{gl}). Unlike to the case of the TeV radius modulus
$\phi$~(\ref{mphi}), its coupling to matter is in general model dependent.
Following the same steps as before, one finds that its coupling to nucleons
relative to gravity $\alpha_{\phi_\perp}$ is given by eq.~(\ref{coupling})
replacing $R$ by $R_\perp$. It follows that $\alpha_{\phi_\perp}$ is strongly
suppressed, except again in the case of $d_\perp=2$, when gauge couplings acquire
logarithmic sensitivity on $R_\perp$. In this case, the resulting scalar force is 
of order ${\cal O}(1)$ relative to gravity --although not exactly computable-- and
therefore in principle within the expected sensitivity of present and future
experiments displayed in fig.~\ref{fig:forces}.

\end{document}